\documentclass[conference]{IEEEtran}

\usepackage{algorithmic}
\usepackage{amsmath,amssymb,amsfonts}
\usepackage{cite}
\usepackage{graphicx}
\usepackage{subcaption}
\usepackage{textcomp}
\usepackage{xcolor}

\renewcommand{\baselinestretch}{0.96}

\begin{document}
\title{Optimizing Validation Phase of Hyperledger Fabric}

\author{
\IEEEauthorblockN{
	Haris Javaid\IEEEauthorrefmark{1} \quad
	Chengchen Hu\IEEEauthorrefmark{1} \quad
	Gordon Brebner\IEEEauthorrefmark{2}
}
\IEEEauthorblockA{
	\IEEEauthorrefmark{1}Xilinx, Singapore \quad 
	\IEEEauthorrefmark{2}Xilinx, USA
}
\IEEEauthorblockA{\{harisj, chengchen, gjb\}@xilinx.com}}

\maketitle

\begin{abstract}
Blockchain technologies are on the rise, and Hyperledger Fabric is one of the most popular permissioned blockchain platforms. In this paper, we re-architect the validation phase of Fabric based on our analysis from fine-grained breakdown of the validation phase's latency. Our optimized validation phase uses a chaincode cache during validation of transactions, initiates state database reads in parallel with validation of transactions, and writes to the ledger and databases in parallel. Our experiments reveal performance improvements of 2$\times$ for CouchDB and 1.3$\times$ for LevelDB. Notably, our optimizations can be adopted in a future release of Hyperledger Fabric. 
\end{abstract}


\section{Introduction}
Blockchain technology is increasingly becoming popular, with applications in various domains such as finance, real estate, supply chains, etc. A blockchain is essentially a distributed ledger of transactions, which is maintained by all the participating nodes of the blockchain network. The transactions represent some business logic and are grouped into blocks which are appended to the ledger. Each node in the network updates its own copy of the ledger with the new block, after consensus is reached amongst the nodes.

In public or permissionless blockchains such as Bitcoin~\cite{Bitcoin} and Ethereum~\cite{Ethereum}, anyone can join the network and the consensus mechanism is based on proof-of-work algorithms which are computationally intensive. In permissioned blochchains, the identity of the nodes is known and authenticated cryptographically. The consensus mechanism is delegated to a few selected nodes in order to reduce bottlenecks in the consensus. Examples include Hyperledger Fabric~\cite{HyperledgerFabric}, Quorum~\cite{Quorum} and Corda~\cite{Corda}. Hyperledger Fabric is one of the most popular platforms as it is open-source and has already been shown to implement many enterprise applications such as food supply chain, healthcare, etc.~\cite{HyperledgerFabricBlog}.

In this paper, we focus on performance improvements for Hyperledger Fabric. The transaction flow in Fabric follows the \textit{execute-order-validate} model, where a transaction is executed first, then ordered into a block, which is finally validated and committed to the ledger. Consequently, some nodes in the Fabric network act as a peer to execute/endorse transactions and validate/commit blocks, while other nodes act as orderers to create new blocks. In addition to the ledger, each peer node uses a state database to keep the global state of the blocks committed so far. Two options are available: (1) LevelDB~\cite{LevelDB} which is an embedded database and allows relatively fast accesses, and (2) CouchDB~\cite{CouchDB} which provides a client-server model and is accessible through REST API over HTTP. Unlike LevelDB, CouchDB allows rich queries over the global state but the accesses are relatively slow.

Many previous performance studies have highlighted validation phase as one of the major bottlenecks~\cite{Baliga2018,Thakkar2018,Gorenflo2019}, in addition to the bottlenecks in consensus mechanism~\cite{Androulaki2018}. For the validation phase, recent optimizations~\cite{Thakkar2018,Gorenflo2019} include validation of transactions in parallel, caching block and identity certificates, and bulk reading from slow CouchDB.

In this paper, we critically examine and thoroughly evaluate the validation phase of Hyperledger Fabric for further improvements. Based on our evaluation and observations, we propose several optimizations around executing various operations in parallel to overlap and hide their latencies. In particular, this paper makes the following contributions:
\begin{itemize}
	\item Validation phase latency is broken down into six components for fine-grained analysis, and extensive experiments are run to highlight the bottlenecks and areas for improvement.
	\item Based on our observations, we find that the validation of transactions is much slower with CouchDB when compared to LevelDB. We propose a chaincode cache to speedup lookups for chaincode information (such as endorsement policy, etc.) instead of always accessing it from the state database during validation of transactions.
	\item We re-architect the validation phase to execute validation of transactions and state database reads in parallel in a way that their subtle dependency is avoided. Furthermore, we execute the ledger writes and state database writes in parallel. Based upon the type of state database used, the writes to history database are also combined with either the ledger write or the state database write operation.
\end{itemize}

We implemented our optimizations in Hyperledger Fabric v1.1, however they are also valid for v1.4. From our experiments with CouchDB, we show that the commit throughput at the peer nodes improved by 2$\times$. For LevelDB, our optimizations improved performance by 1.3$\times$. Most importantly, the proposed optimizations can be adopted in a future release of Hyperledger Fabric.

The rest of the paper is organized as follows. Section~\ref{sec:fabric_arch} provides an overview of Fabric with details of its validation phase. Section~\ref{sec:optimizations} presents our evaluation methodology and in-depth analysis of the validation phase, along with our optimized validation phase. Experimental results of our optimizations are discussed in Section~\ref{sec:experiments}. Section~\ref{sec:related_work} describes the related work, and the paper is concluded in Section~\ref{sec:conclusion}.

\section{Hyperledger Fabric Architecture}\label{sec:fabric_arch}

\subsection{Overview}
Hyperledger Fabric is an open-source, enterprise-grade implementation platform for permissioned blockchains. A Fabric network consists of different types of nodes, such as peers, orderers, clients, etc., where each node has an identity provided by the \textit{membership service provider}.

An endorsing peer both executes/endorses transactions and validates/commits blocks to the ledger. A non-endorsing peer only validates/commits blocks to the ledger. Execution of transactions is enabled by smart contracts or chaincodes, which represent the business logic and are instantiated on the endorsing peers.

The ordering sevice consists of orderers which use a consensus mechanism to establish a total order for the transactions. A block is created from the ordered transactions, and then broadcast to the peers. Multiple pluggable consensus mechanisms are available, such as Apache Kafka/Zookeper~\cite{Kafka} or Raft~\cite{Ongaro2014} based consensus mechanism.

\subsection{Transaction Flow}
\begin{figure}[t]
	\centering
	\includegraphics[width=0.95\columnwidth]{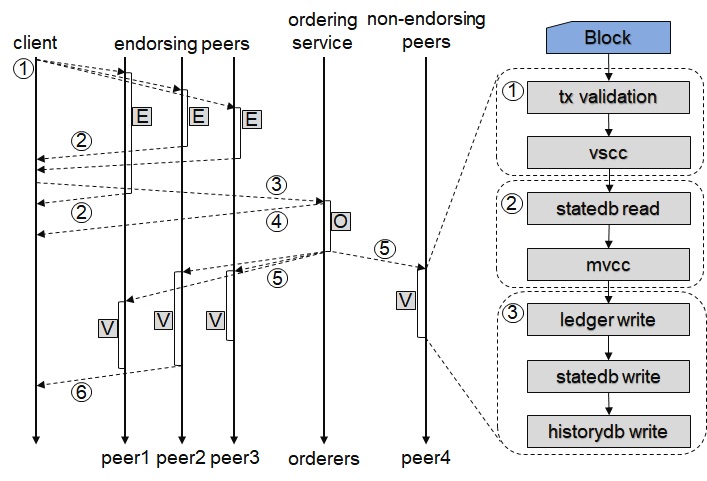}
	\caption{Transaction Flow in Hyperledger Fabric.}
	\label{fig:transaction_flow}
\vspace{-0.5cm}
\end{figure}

A transaction flows through the various nodes of a Fabric network as illustrated in Figure~\ref{fig:transaction_flow} (left-hand side, see~\cite{Androulaki2018} for more details). A client creates a transaction and sends it to a number of endorsing peers (step 1). Each endorsing peer executes the transaction against its own state database, to compute the read-write set of the transaction (marked as \textit{E}). The read set is the keys accessed and their version numbers, while the write set is the keys to be updated with their new values. If there are no errors during the execution of the transaction, the peer sends back an endorsement to the client (step 2). After the client has gathered enough endorsements, it submits the transaction with its endorsements to the ordering service (step 3).

The ordering service responds back to the client after the transaction has been accepted for inclusion into a block (step 4). A block is created from the ordered transactions when either a user-configured timeout has expired or user-configured limit on block size has reached. Once a block is created (marked as \textit{O}), the orderer broadcasts it to all the peers (step 5). Each peer validates all the transactions of the block and then commits it to the ledger and state database (marked as \textit{V}). Finally, one of the peers sends a notification to the client that the transaction has been committed (step 6).

Figure~\ref{fig:transaction_flow} shows the operations of the validation phase in more detail on the right-hand side. On receiving the block from the orderer (or another peer) through the Gossip protocol, a peer checks the syntactic structure of the block, and then sends it through a pipeline of various operations. In step 1, each transaction in the block is syntactically validated. Then, \textit{vscc} (validation system chaincode) is run on each transaction where the endorsements are validated and the endorsement policy of the associated chaincode is evaluated. A transaction is marked as invalid if its endorsement policy is not satisfied.

In step 2, \textit{mvcc} (multi-version concurrency control) check is applied. This check ensures that there are no read-write conflicts between the valid transactions; in other words, it avoids the double-spending problem. The read set of each transaction is computed again by accessing the state database, and is compared to the read set from the endorsement phase. If these read sets are different, then some other transaction (either in this block or an earlier block) has already modified the same keys, and hence this transaction is marked as invalid. For LevelDB, the state database read operation is integrated with the mvcc operation. However, for CouchDB, all the keys for all the transactions are read in a bulk operation before starting the \textit{mvcc} operation.

In the final step 3, the block is committed. First, the entire block is written to the ledger with its transactions' valid/invalid flags. Then, the write sets of the valid transactions are committed to the state database. Finally, the history database is updated to keep track of which keys have been modified by which blocks and transactions.

\section{Evaluation Methodology And Optimizations}\label{sec:optimizations}

\subsection{Fabric Network Setup}
We created a Fabric network with two organizations, where each organization had two endorsing peers and a certificate authority. We used Kafka based ordering service with two orderer nodes, four Kafka brokers and three Zookeeper nodes. Each peer is run on a virtual machine which is allocated 16 Intel Xeon 4416 @ 2.1GHz vCPUs with 32GB RAM, 50GB hard disk, and configured with Ubuntu 16.04 LTS. All the machines were connected through a 1Gbps network.

\subsection{Application}
We used the \textit{smallbank} benchmark from Hyperledger Caliper~\cite{HyperledgerCaliper} to test our Fabric network. The \textit{smallbank} benchmark is representative of a banking application, where its chaincode implements functions such as creation of a user account, transfer money, deposit cash, etc. For each experiment, the clients were configured to create random transactions from the pool of available functions. A total of 30,000 transactions were created and sent to the peers at the rate closer to their saturation point~\cite{Thakkar2018} (i.e., peer throughput is stable, which we determined empirically through our experiments). A single channel was created on the peers with the endorsement policy of at least one signature from each organization. We used a virtual machine with 8 vCPUs to run the \textit{smallbank} clients.

\subsection{Metrics}\label{sec:metrics}
Since our goal is to critically examine the validation phase of Fabric, we use the \textit{commit throughput} and \textit{block validation latency} as the primary metrics for performance evaluation. Commit throughput is defined as the rate at which transactions are committed to the ledger by the peer. The block validation latency is the total time taken by the peer to validate and commit the entire block. Unlike previous works which consider only the coarse-grained latencies at the validation, \textit{mvcc}, and commit operations~\cite{Thakkar2018}, we breakdown the validation latency into six components for fine-grained analysis:

\begin{itemize}
	\item \textit{vscc}: time spent in syntactic validation of the transactions as well as the execution of \textit{vscc} for validation of endorsement policy.
	\item \textit{statedb\_read}: time spent in reading from the state database. For LevelDB, this is always zero as the state database reads are integrated with the \textit{mvcc} step.
	\item \textit{mvcc}: time spent in executing the \textit{mvcc} checks. For LevelDB, this latency also includes the time spent in reading from the state database.
	\item \textit{ledger\_write}: time spent in committing the block to the ledger.
	\item \textit{statedb\_write}: time spent in committing the write sets of transactions to state database.
	\item \textit{others}: time spent in miscellaneous operations.
\end{itemize}

We did not use the Caliper tool to measure the above metrics because (1) It had issues missing block events at higher transaction rates (also reported in~\cite{Baliga2018,Sharma2018}), and (2) It can only measure throughput and latency at the client level which does not provide in-depth insights of the validation phase. We instrumented the Fabric code to log timestamps at various points through the validation phase, and then calculated the above metrics after an experiment has finished. Each experiment was repeated 20 times to compute average metrics.

\subsection{Anlaysis of Validation Phase}
We run Fabric v1.1 using the setup described above as our baseline for analysis. We changed the number of \textit{vscc} threads and the block size, which are the two most important configuration parameters of the validation phase~\cite{Baliga2018,Thakkar2018}.

\subsubsection{LevelDB}
\begin{figure}[t]
	\centering
	\includegraphics[width=0.95\columnwidth]{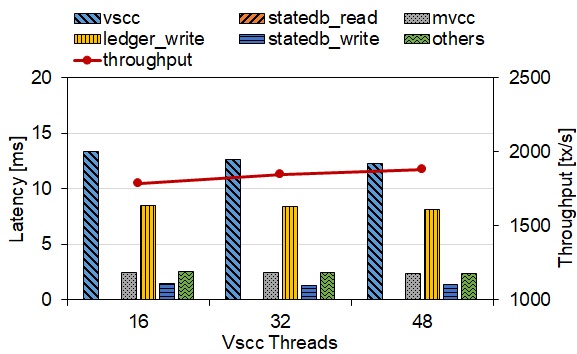}
	\caption{Original validation phase: LevelDB vs. vscc threads.}
	\label{fig:vscc_leveldb_seq}
\vspace{-0.25cm}
\end{figure}

Figure~\ref{fig:vscc_leveldb_seq} shows the breakdown of validation latency and commit throughput for LevelDB with a block size of 50, and varying \textit{vscc} trheads form 16 to 48. We make the following observations here:
\begin{itemize}
	\item The \textit{vscc} operation even with multiple threads is still the bottleneck. The \textit{vscc} latency reduces with an increase in the number of threads; however, the overall throughput only improves slightly because all the other latencies do not improve with \textit{vscc} threads.
	\item The block commit is dominated by the writes to the ledger instead of the writes to the state database because LevelDB provides relatively fast accesses.
	\item The \textit{others} latency is not negligible and in fact, it is comparable to \textit{mvcc} and more than the \textit{statedb\_write} latency. This is because it is dominated by the time spent in writing to the history database.
\end{itemize}

\begin{figure}[t]
	\centering
	\includegraphics[width=0.95\columnwidth]{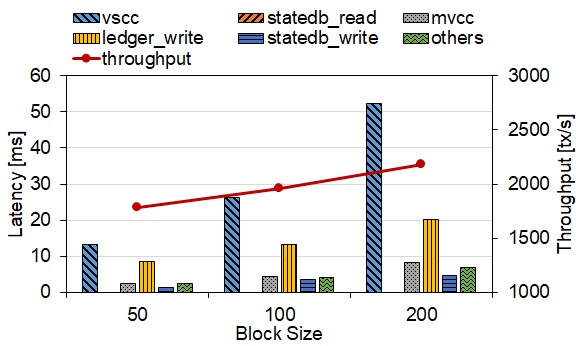}
	\caption{Original validation phase: LevelDB vs. block size.}
	\label{fig:bs_leveldb_seq}
\vspace{-0.5cm}
\end{figure}

The results from varying the block size from 50 to 200 are reported in Figure~\ref{fig:bs_leveldb_seq}, and our earlier observations still hold. The interesting point to note here is that the improvement in throughput is more noticeable because the cost of committing the block to the ledger and state database is better amortized for larger blocks.

\begin{figure}[t]
	\centering
	\includegraphics[width=0.95\columnwidth]{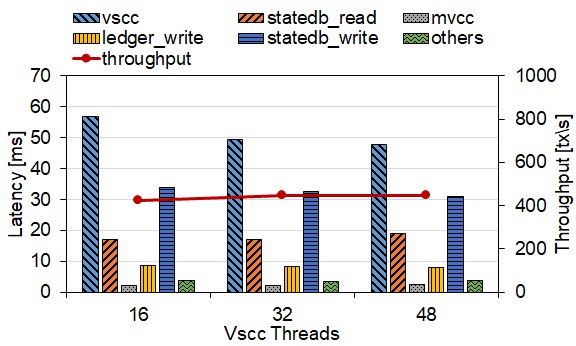}
	\caption{Original validation phase: CouchDB vs. vscc threads.}
	\label{fig:vscc_couchdb_seq}
\vspace{-0.25cm}
\end{figure}

\begin{figure}[t]
	\centering
	\includegraphics[width=0.97\columnwidth]{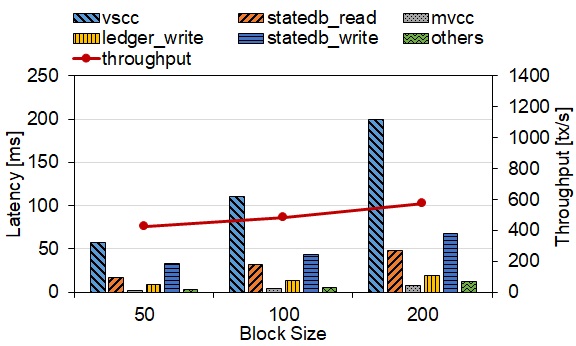}
	\caption{Original validation phase: CouchDB vs. block size.}
	\label{fig:bs_couchdb_seq}
\vspace{-0.5cm}
\end{figure}

\subsubsection{CouchDB}
The results for CouchDB with varying \textit{vscc} threads are presented in Figure~\ref{fig:vscc_couchdb_seq}. The distribution of the latencies is quite different compared to LevelDB. The noteworthy observations here are:
\begin{itemize}
	\item The \textit{vscc} latency is the bottleneck, and is almost 4$\times$ that of the \textit{vscc} latency when LevelDB is used. This comes as a surprise since the \textit{vscc} operation should be independent of the type of state database. It turns out that the chaincode information such as version, endorsement policy, etc. is stored in the state database. Since \textit{vscc} operation enforces endorsement policy, for each transaction, it accesses the state database to retrieve this information. Given that CouchDB is accessed through REST API which is slow, the \textit{vscc} latency is much more when compared to LevelDB.
	\item It is the state database accesses where most of the time is spent. The \textit{statedb\_read} latency is significantly high. Furthermore, the block commit is dominated by the \textit{statedb\_write} latency instead of the \textit{ledger\_write} in contrast to LevelDB.
	\item The \textit{mvcc} and \textit{others} latencies are not significant, and are more or less the same as LevelDB.
\end{itemize}

The latency and throughput trends are similar when the block size is changed from 50 to 200, which are shown in Figure~\ref{fig:bs_couchdb_seq}.

\subsection{Optimized Validation Phase}
The observations from our analysis can be summarized into two main points: (1) Only the \text{vscc} operation benefits from multiple CPUs as it uses multiple threads to validate transactions in parallel. The other operations of the validation phase are sequential, and do not fully utilize the available computational resources, and (2) Accessing state database for information that does not change very often can have a significant negative impact on the performance.

\begin{figure}[t]
	\centering
	\includegraphics[width=.67\columnwidth]{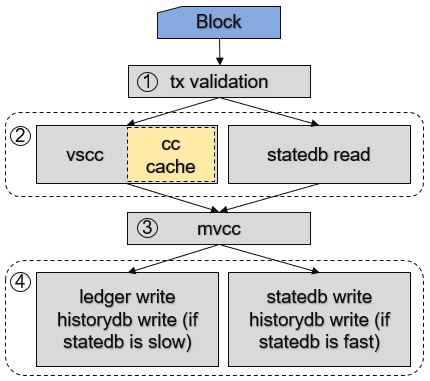}
	\caption{Optimized Validation Phase for Hyperledger Fabric.}
	\label{fig:opt_validation_phase}
\vspace{-0.5cm}
\end{figure}

We re-architect the validation phase by executing as many operations of the validation phase in parallel as possible and use a cache for chaincode information. Our optimized validation phase is depicted in Figure~\ref{fig:opt_validation_phase}. Like the original validation phase, in step 1, all the transactions are syntactically validated. If a transaction is ill-formed, then its flag is set to invalid. Afterwards, in step 2, we initiate both the \textit{vscc} and the state database read operations in parallel. Recall that the state database read operation here is only applicable to CouchDB, and is ignored for LevelDB. The motivation is that the latencies of these two operations can be overlapped with each other to reduce the overall block validation latency (see Figure~\ref{fig:vscc_couchdb_seq}).

The \textit{vscc} operation is modified to leverage a chaincode cache (marked as \textit{cc cache}) implemented as a map of chaincode id to its detailed information such as chaincode name, version, endorsement policy, etc. Typically, only a few tens of chaincodes will be instantiated, so their information can be cached for fast accesses by avoiding state database reads. During validation of each transaction, its associated chaincode is searched in the cache. If the cache lookup results in a miss, then state database is accessed to retrieve the chaincode information and the cache is updated. When a chaincode is upgraded, its entry from the cache is deleted so that the new information can be retrieved from the state database again. Another possibility is to clear the chaincode cache at the start of each block (instead of clearing cache at chaincode upgrade). However, for this implementation, the block size should be large enough to amortize the cost of clearing and re-populating the cache for every block.

The state database read operation computes the read sets of all the syntactically valid transactions by reading from the state database. It is possible that some of these syntactically valid transactions might be marked as invalid by the \textit{vscc} operation. This is addressed in step 3, where the \textit{mvcc} checks are only applied to transactions that have been validated by the \textit{vscc} operation, discarding all other invalid transactions. In other words, the \textit{mvcc} operation discards the invalid transactions by ignoring their read sets computed during the state database read operation. The output of the \textit{mvcc} operation will be exactly the same as the original validation phase.

In the final step 4, the block is committed to the ledger and the two databases in parallel. First, the ledger write and state database write operations are executed in parallel to hide the latency of the faster operation (see Figures~\ref{fig:vscc_leveldb_seq} and~\ref{fig:vscc_couchdb_seq}). Then, for LevelDB (fast database), the write to history database is initiated after the state database write operation. For CouchDB (slow database), history database write operation is combined with the ledger write operation instead.

It is possible that the write to ledger succeeds while the state (or history) database write operation fails, then the peer raises a panic error following the existing recovery mechanism~\cite{Nathan2018}. On restart, the peer fetches the failed/missed blocks from other peers to reconstruct the ledger and the databases. In our optimized validation phase, it is also possible that the writes to either one or both the databases succeeds while the ledger write operation fails, then the databases will be inconsistent with the ledger. In this scenario, we propose to retry ledger write for a number of times before reporting a panic error. As explained above, the peer will reconstruct the ledger and databases on restart. Alternatively, the databases could be rolled back by reverting to previous version/snapshot or re-writing the old values. Note that an in-depth study of crash/fault tolerance of a peer in case of write failures will be addressed in a future work.

\section{Experimental Results}\label{sec:experiments}

\subsection{Implementation}
We used the same setup as described in Section~\ref{sec:optimizations} to evaluate the optimized validation phase. In our implementation of the optimized validation phase, we did not separate the syntactic validation from \textit{vscc} validation due to the significant code refactor effort required. This did not affect our experiments as all the transactions were well-formed. Likewise, we did not implement the cache update mechanism when a chaincode is upgraded as our experiments used the same chaincode during the entire run. However, it should be noted that the Fabric code refactor proposal in~\cite{JiraFab12221} for the next release will greatly simplify the implementation of our proposals.

All the parallel operations were implemented as goroutines. We added two more latencies, in addition to the ones described in Section~\ref{sec:metrics}:
\begin{itemize}
	\item \textit{vscc\_statedb\_read}: total time spent in syntactic and \textit{vscc} validation of transactions, and reading from the state database when these operations are executed in parallel.
	\item \textit{ledger\_statedb\_write}: time spent in committing the block to the ledger, and state and history databases when these operations are executed in parallel.
\end{itemize}

\subsection{Results}
\subsubsection{LevelDB}
\begin{figure}[t]
	\centering
	\includegraphics[width=0.95\columnwidth]{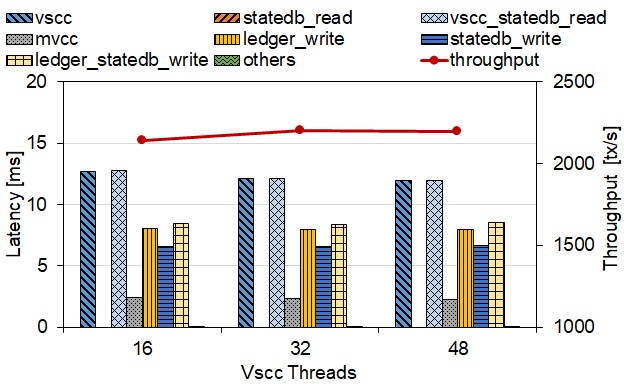}
	\caption{Optimized validation phase: LevelDB vs. vscc threads.}
	\label{fig:vscc_leveldb_opt}
\vspace{-0.25cm}
\end{figure}

The results for LevelDB with block size of 50 and varying \textit{vscc} threads are reported in Figure~\ref{fig:vscc_leveldb_opt}. As expected, the latencies for \textit{vscc}, state database read and \textit{mvcc} operations are similar to the original validation phase (see Figure~\ref{fig:vscc_leveldb_seq}). Furthermore, the \textit{vscc\_statedb\_read} latency is more or less the same as the \textit{vscc} latency, which means the overheads of parallel execution are negligible.

More importantly, the \textit{statedb\_write} latency is completely hidden by the \textit{ledger\_write} latency. The \textit{statedb\_write} latency in the optimized validation phase is more than the original phase because it includes the time taken to write to the history database as well. As a result, the \textit{others} latency is now almost zero. For 16 \textit{vscc} threads, the throughput increased from 1,784 to 2,124 transactions/second which is a 1.2$\times$ improvement.

\begin{figure}[t]
	\centering
	\includegraphics[width=0.95\columnwidth]{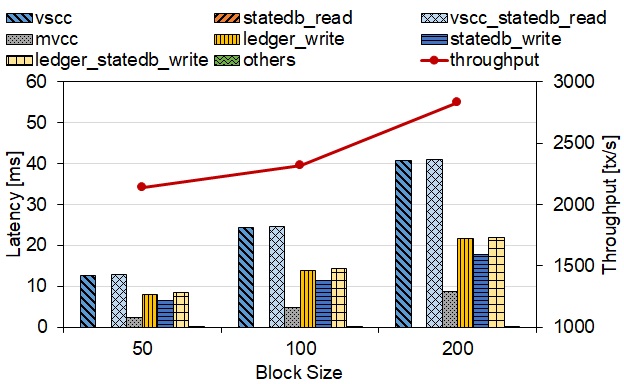}
	\caption{Optimized validation phase: LevelDB vs. block size.}
	\label{fig:bs_leveldb_opt}
\vspace{-0.5cm}
\end{figure}

The validation latency and commit throughput for varying block sizes are presented in Figure~\ref{fig:bs_leveldb_opt}. The benefits of parallel execution are evident again as the \textit{ledger\_statedb\_write} latency is about 30\% lower than that of the sum of \textit{ledger\_write} + \textit{statedb\_write} + \textit{others} latencies from Figure~\ref{fig:bs_leveldb_seq} (8.5 ms compared to 12.5 ms for block size of 50). For larger block sizes, the throughput improvement is better (1.3$\times$ for 200 compared to 1.2$\times$ for 50) because more transactions result in chaincode cache hits than misses. To summarize, our optimized validation phase achieves a 1.3$\times$ improvement in throughput (2,835 compared to 2,178 transactions/second) with 16 \textit{vscc} threads and block size of 200.

\begin{figure}[t]
	\centering
	\includegraphics[width=0.95\columnwidth]{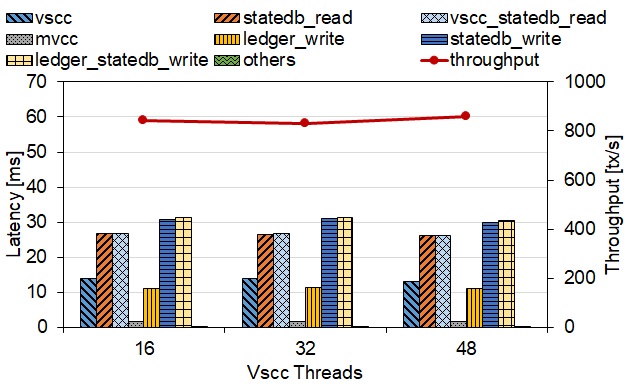}
	\caption{Optimized validation phase: CouchDB vs. vscc threads.}
	\label{fig:vscc_couchdb_opt}
\vspace{-0.25cm}
\end{figure}

\begin{figure}[t]
	\centering
	\includegraphics[width=0.99\columnwidth]{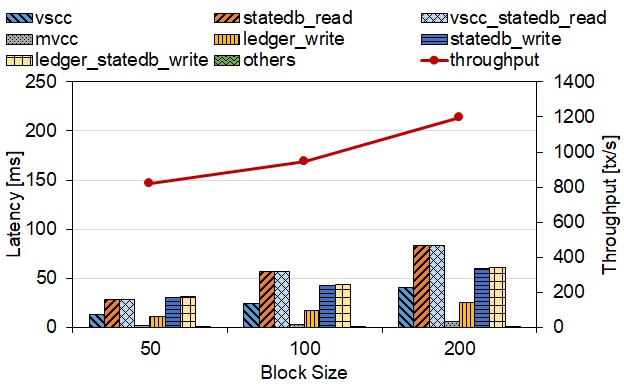}
	\caption{Optimized validation phase: CouchDB vs. block size.}
	\label{fig:bs_couchdb_opt}
\vspace{-0.5cm}
\end{figure}

\subsubsection{CouchDB}
The CouchDB results are even more interesting, and are reported in Figure~\ref{fig:vscc_couchdb_opt} for block size of 50 and varying \textit{vscc} threads. First, the \textit{vscc} latency decreased significantly, by about 4$\times$, from 57 ms in Figure~\ref{fig:vscc_couchdb_seq} to about 14 ms. This is due to the chaincode cache we introduced in Section~\ref{sec:optimizations}, which enables the \textit{vscc} operation with CouchDB to perform on par with LevelDB. Furthermore, the \textit{statedb\_read} latency completely hides the \textit{vscc} latency, making the \textit{vscc\_statedb\_read} latency more or less the same as \textit{statedb\_read} latency. An interesting point to note here is that the \textit{statedb\_read} latency is more than its counterpart in Figure~\ref{fig:vscc_couchdb_seq}. The reason is that both the \textit{vscc} and state database read operations are computationally intensive and compete for the available CPU resources. However, the overall \textit{vscc\_statedb\_read} latency is still less than the sum of \textit{vscc} and \textit{statedb\_read} latencies from Figure~\ref{fig:vscc_couchdb_seq}.

On the block commit operations, the \textit{ledger\_write} latency is completely hidden by the \textit{statedb\_write} latency. The \textit{ledger\_write} latency in this case also includes the time taken to write to the history database, and hence is more than the \textit{ledger\_write} latency in Figure~\ref{fig:vscc_couchdb_seq}. Therefore, the \textit{others} latency is almost negligible. Overall, the throughput improvement is about 2$\times$ for 16 \textit{vscc} threads, where the throughput increased from 424 to 841 transactions/second.

Similar improvements are observed across varying block sizes as reported in Figure~\ref{fig:bs_couchdb_opt}. Most importantly, our optimizations can achieve a commit throughput of about 1,200 transactions/second with a block size of 200. Most of this speedup resulted from the reduction in \textit{vscc} latency and its parallel execution with state database read (3$\times$ compared to Figure~\ref{fig:bs_couchdb_seq}) and overlapped writes during block commit (1.6$\times$ compared to Figure~\ref{fig:bs_couchdb_seq}). For perspective, the best throughput reported with CouchDB in literature~\cite{Thakkar2018} is 700 transactions/second with a block size of 500 transactions and peers with 32 vCPUs (in our setup peers run on 16 vCPU machines).

\subsection{Remarks}
Based on our results, we make the following remarks for the optimized validation phase:
\begin{itemize}
	\item The chaincode cache in \textit{vscc} operation benefits the most when CouchDB is the state database and higher block sizes are used. The expensive CouchDB accesses are avoided due to the cache hits. Since LevelDB itself is relatively fast, the chaincode cache impact is more noticeable with LevelDB when higher block sizes are used as more database lookups are served from the cache. We believe that such a cache could be generalized to implement a generic caching layer for the state database~\cite{JiraFab103}.
	\item The parallel execution of vscc and state database read operations again benefit CouchDB. In general, this optimization is geared towards state databases that are slow and allow bulk read option.
	\item The parallel execution of writes to ledger, state database and history database during block commit benefits both LevelDB and CouchDB.	This is because either the ledger write or the state database write latency can hide the other two latencies depending on whether the state database is slow or fast.
\end{itemize}

\section{Related Work}\label{sec:related_work}
Hyperledger Fabric is a recent platform, and thus has been evolving very rapidly. Here, we survey the most relevant work on benchmarking and improving peer performance. For performance bottlenecks in consensus, readers are referred to~\cite{Androulaki2018, Bano2017, Vukolic2016}.

The detailed architecture of Fabric v1.0 was published in~\cite{Androulaki2018}, where the authors explained their design choices compared to v0.6, and demonstrated the new architecture's performance and scalability using a cryptocurrency application. Dinh et al.~\cite{Dinh2017} proposed a framework to evaluate private blockchains and compared Fabric v0.6, Ethereum and Parity. Likewise, the authors in~\cite{Pongnumkul2017} compared the performance of Fabric v0.6 and Ethereum. Nasir et al.~\cite{Nasir2018} compared the performance of Fabric v0.6 and v1.0. The authors in~\cite{Baliga2018} evaluated Fabric v1.0 to show that application level parameters such as transaction size, chaincode size, etc. significantly impact performance. Our work differs from these works as they only evaluate the Fabric platform without proposing any architectural changes or optimizations.

Our work is closest to~\cite{Thakkar2018, Gorenflo2019, Sharma2018}. Thakkar et al.~\cite{Thakkar2018} performed an extensive study of the Fabric v1.0, and found that the major bottlenecks are in repeated deserialization of indentities/certificates, sequential validation of transactions during \textit{vscc}, and slow CouchDB accesses. They introduced caching identities, parallel \textit{vscc} and bulk read for CouchDB, which were incorporated in the v1.1 release. Hence, our optimizations are on top of their proposals, and we go one step further by caching chaincode information and running operations other than \textit{vscc} in parallel as well.

In a recent work~\cite{Gorenflo2019}, Gorenflo et al. focused on re-architecting both the orderer and peer in Fabric v1.2. For orderers, they proposed to use only the transaction ids instead of full transactions. For peers, they parallelized \textit{vscc} by running it across multiple blocks, cached unmarshalled blocks, and used an in-memory hash table instead of a state database. The authors in~\cite{Sharma2018} proposed re-ordering transactions and early abort mechanism in the orderers to minimize conflicting transactions in a block. Like these works, we also propose architectural changes, however, our optimizations are focused on improving other operations (such as chaincode accesses, and database reads and writes in parallel) which have not been explored before, and can be combined with these existing works. For example, assume that the in-memory hash table from~\cite{Gorenflo2019} acts as a caching layer for state database~\cite{JiraFab103}, then the parallel reads and writes from state database can be combined with~\cite{Gorenflo2019} to further improve Fabric's performance.


\section{Conclusion}\label{sec:conclusion}
In this paper, we conduct a fine-grained evaluation of the validation phase of Hyperledger Fabric, and based on our analysis, we re-architect the validation phase. Our optimized validation phase uses a chaincode cache during validation of transactions, executes state database reads in parallel with validation of transactions, as well as writes to the ledger and databases in parallel. We implemented our optimizations in Fabric v1.1 (also valid for v1.4), and show that for CouchDB the throughput improves by 2$\times$ (from 575 to 1,196 transactions/second), while the improvement with LevelDB is 1.3$\times$ (from 2,178 to 2,835 transactions/second). Our optimizations are orthogonal to previous works~\cite{Thakkar2018,Gorenflo2019}, and thus can be adopted in a future release of  Fabric.


\begin{thebibliography}{10}

\bibitem{CouchDB}
{Apache CouchDB}.
\newblock http://couchdb.apache.org/.

\bibitem{Kafka}
{Apache Kafka}.
\newblock http://kafka.apache.org/.

\bibitem{Bitcoin}
{Bitcoin}.
\newblock https://bitcoin.org/en/.

\bibitem{Corda}
{Corda}.
\newblock https://www.corda.net.

\bibitem{Ethereum}
{Ethereum}.
\newblock https://ethereum.org.

\bibitem{HyperledgerCaliper}
{Hyperledger Caliper}.
\newblock https://www.hyperledger.org/projects/caliper.

\bibitem{HyperledgerFabric}
{Hyperledger Fabric}.
\newblock https://www.hyperledger.org/projects/fabric.

\bibitem{LevelDB}
{LevelDB in Go}.
\newblock https://github.com/syndtr/goleveldb/.

\bibitem{Quorum}
{Quorum}.
\newblock https://www.jpmorgan.com/global/Quorum.

\bibitem{Androulaki2018}
Elli Androulaki, Artem Barger, Vita Bortnikov, Christian Cachin, Konstantinos
  Christidis, Angelo {De Caro}, David Enyeart, Christopher Ferris, Gennady
  Laventman, Yacov Manevich, Srinivasan Muralidharan, Chet Murthy, Binh Nguyen,
  Manish Sethi, Gari Singh, Keith Smith, Alessandro Sorniotti, Chrysoula
  Stathakopoulou, Marko Vukoli{\'{c}}, Sharon~Weed Cocco, and Jason Yellick.
\newblock {Hyperledger Fabric: A Distributed Operating System for Permissioned
  Blockchains}.
\newblock In {\em EuroSys}, 2018.

\bibitem{Baliga2018}
Arati Baliga, Nitesh Solanki, Shubham Verekar, Amol Pednekar, Pandurang Kamat,
  and Siddhartha Chatterjee.
\newblock {Performance Characterization of Hyperledger Fabric}.
\newblock In {\em Crypto Valley Conference on Blockchain Technology (CVCBT)},
  2018.

\bibitem{Bano2017}
Shehar Bano, Alberto Sonnino, Mustafa Al-Bassam, Sarah Azouvi, Patrick McCorry,
  Sarah Meiklejohn, and George Danezis.
\newblock {Consensus in the Age of Blockchains}.
\newblock In {\em CoRR, arXiv:1711.03936}, 2017.

\bibitem{HyperledgerFabricBlog}
Hyperledger Blog.
\newblock {Forbes Blockchain 50: Half of the biggest companies deploying
  blockchain use Hyperledger}.
\newblock https://www.hyperledger.org/blog/2019/04/18/{\_}{\_}trashed, 2019.

\bibitem{Dinh2017}
Tien Tuan~Anh Dinh, Ji~Wang, Gang Chen, Rui Liu, Beng~Chin Ooi, and Kian-Lee
  Tan.
\newblock {BLOCKBENCH: A Framework for Analyzing Private Blockchains}.
\newblock In {\em Proceedings of the 2017 ACM International Conference on
  Management of Data (SIGMOD)}, 2017.

\bibitem{Gorenflo2019}
Christian Gorenflo, Stephen Lee, Lukasz Golab, and S.~Keshav.
\newblock {FastFabric: Scaling Hyperledger Fabric to 20,000 Transactions per
  Second}.
\newblock In {\em CoRR, arXiv:1901.00910}, 2019.

\bibitem{JiraFab103}
Hyperledger~Fabric JIRA.
\newblock {FAB-103 Cache of the World State for Improved Performance}.
\newblock https://jira.hyperledger.org/browse/FAB-103.

\bibitem{JiraFab12221}
Hyperledger~Fabric JIRA.
\newblock {FAB-12221 Validator/Committer Refactor}.
\newblock https://jira.hyperledger.org/browse/FAB-12221?filter=12526.

\bibitem{Nasir2018}
Qassim Nasir, Ilham~A. Qasse, Manar {Abu Talib}, and Ali~Bou Nassif.
\newblock {Performance analysis of hyperledger fabric platforms}.
\newblock {\em Security and Communication Networks}, 2018.

\bibitem{Nathan2018}
Senthil Nathan.
\newblock {Failure and Recovery of StateDB in Hyperledger Fabric v1.1}.
\newblock
  https://blockchain-fabric.blogspot.com/2018/05/failure-and-recovery-of-statedb-in.html,
  2018.

\bibitem{Ongaro2014}
Diego Ongaro and John Ousterhout.
\newblock {In Search of an Understandable Consensus Algorithm}.
\newblock In {\em Proceedings of the 2014 USENIX Conference on USENIX Annual
  Technical Conference (ATC)}, pages 305--320, 2014.

\bibitem{Pongnumkul2017}
Suporn Pongnumkul, Chaiyaphum Siripanpornchana, and {Suttipong Thajchayapong}.
\newblock {Performance Analysis of Private Blockchain Platforms in Varying
  Workloads}.
\newblock In {\em International Conference on Computer Communication and
  Networks (ICCCN)}, 2017.

\bibitem{Sharma2018}
Ankur Sharma, Felix~Martin Schuhknecht, Divya Agrawal, and Jens Dittrich.
\newblock {How to Databasify a Blockchain: the Case of Hyperledger Fabric}.
\newblock In {\em CoRR, arXiv:1810.13177}, 2018.

\bibitem{Thakkar2018}
Parth Thakkar, Senthil Nathan, and Balaji Vishwanathan.
\newblock {Performance Benchmarking and Optimizing Hyperledger Fabric
  Blockchain Platform}.
\newblock In {\em 26th IEEE International Symposium on the Modeling, Analysis,
  and Simulation of Computer and Telecommunication Systems (MASCOTS)}, 2018.

\bibitem{Vukolic2016}
Marko Vukoli{\'{c}}.
\newblock {The quest for scalable blockchain fabric: Proof-of-work vs. BFT
  replication}.
\newblock {\em Lecture Notes in Computer Science (including subseries Lecture
  Notes in Artificial Intelligence and Lecture Notes in Bioinformatics)}, 2016.

\end{thebibliography}

\end{document}